\documentclass[11pt,usletter]{article}

\usepackage{amsthm}
\usepackage{jheppub}
\pdfoutput=1

\usepackage{graphicx}
\usepackage{hyperref}
\usepackage{color}
\usepackage{epsfig}
\usepackage{amsfonts}
\usepackage{amssymb}
\usepackage{physics}
\usepackage{amsmath}
\usepackage{amsthm}
\usepackage{subfig}
\usepackage{bm}
\usepackage{mathrsfs}
\usepackage{bbm}
\usepackage{cancel}
\usepackage{accents}

\def\ba{\begin{align}}
\def\ea{\end{align}}
\def\be{\begin{equation}}
\def\ee{\end{equation}}
\def\bea{\begin{eqnarray}}
\def\eea{\end{eqnarray}}
\usepackage[utf8]{inputenc}
\usepackage[english]{babel}

\newtheorem{theorem}{Theorem}

\newtheorem{algorithm}[theorem]{Algorithm}

\begin{document}

\title{Quantum algorithms for conformal bootstrap}

\date{\today}
\author{Ning Bao$^{\ket{\psi}}$}
\author{Junyu Liu$^{\Delta,J}$}

\affiliation{$^{\ket{\psi}}$Berkeley Center for Theoretical Physics,\\ University of California, Berkeley, CA 94720, USA}
\affiliation{$^\Delta$Walter Burke Institute for Theoretical Physics,\\ California Institute of Technology, Pasadena, CA 91125, USA}
\affiliation{$^J$Institute for Quantum Information and Matter,\\ California Institute of Technology, Pasadena, CA 91125, USA}

\emailAdd{ningbao75@gmail.com}
\emailAdd{jliu2@caltech.edu}

\abstract{
With the help of recent developments in quantum algorithms for semidefinite programming, we discuss the possibility for quantum speedup for the numerical conformal bootstrap in conformal field theory. We show that quantum algorithms may have significant improvement in the computational performance for several numerical bootstrap problems. 
}


\maketitle
\flushbottom

\section{Introduction}
Quantum computing is a rapidly growing area both theoretically and experimentally. In the near future, people will be able to manipulate $\mathcal{O}(50)\sim \mathcal{O}(100)$ qubits, a regime which could potentially run algorithms that can outperform classical computers, even with quite high levels of noise. The coming era, the so-called Noisy Intermediate-Scale Quantum(NISQ) era, may, therefore, allow physicists to simulate quantum physics and solve physics problems too difficult for classical computers \cite{Qu}. 
\\
\\
One of the earliest motivation for designing quantum computers was the simulation of quantum field theory \cite{Feynman:1981tf}. With infinitely many degrees of freedom and complicated mathematical structures, this is a problem that has traditionally been difficult for classical computing. Some interesting work has been pursued as to whether quantum algorithms, by contrast, could be designed to simulate field theory processes, such as time evolution of wave packets, and scattering problems in high energy theory (for example, see \cite{Jordan:2011ne,Jordan:2011ci,Jordan:2014tma,Jordan:2017lea,JP}). Indeed, foundational aspects of quantum algorithmic design (such as complexity theory) could correspond to deep concepts in quantum field theory, holography or quantum gravity (for example, see
\cite{Brown:2015bva,Brown:2015lvg,Brown:2016wib}).
\\
\\
For special classes of quantum field theories, such as theories with conformal symmetry, non-perturbative technologies are widely used, that sometimes allow for a solution to the theory through either analytical or numerical techniques. A leading example of this prevalent in the current high energy theory community is called the conformal bootstrap. The conformal bootstrap is an approach that extracts physical data, such as dimensions of states and Operator Product Expansion (OPE) coefficients, from internal non-trivial consistency relations called crossing equations derived from conformal symmetry. Currently, the conformal bootstrap is widely used in conformal field theory (CFT), string theory and statistical physics (for example, see \cite{El-Showk:2014dwa,Kos:2014bka,Simmons-Duffin:2015qma,Kos:2015mba,Iliesiu:2015qra,Lin:2015wcg,Poland:2016chs,Poland:2018epd}). One of the most important developments in this field has been the advent of the numerical bootstrap as a method for finding high precision numerical solutions for critical components in the three dimensional Ising model, an old but interesting puzzle in statistical physics \cite{El-Showk:2014dwa}.
\\
\\
For completeness, we review the basic form of the conformal bootstrap problem. In its simplest case, the crossing equation, derived from the crossing symmetry of the conformal four-point function, can be written in the following form
\begin{align}
{F_{{\Delta _0},{J_0}}}(x) + \sum\limits_{\Delta ,J} {\lambda _{\Delta ,J}^2{F_{\Delta ,J}}(x)}  = 0
\end{align}
where the $x$'s are particular combinations of spacetime points, $F$ is a known function, and the summation over $\Delta$ and $J$ (corresponding to the spectrum of the operator appearing in the OPE) are not known. Alternatively, we could also have known operators in the spectrum. Typically, however, we only know the identity operator, $\Delta_0$ and $J_0$.
\\
\\
The goal is then to solve the spectrum from such a crossing equation. In general, the function $F$ is complicated enough such that determining the spectrum analytically is intractable. Despite this, however, numerical methods have proven successful for this problem. The particular method used here is based on semidefinite programming, a widely used method for operational research and related areas. We can begin by writing a generic form of the linear functional
\begin{align}
\alpha  = \sum\limits_m {{\alpha _m}\partial _x^m{|_{x = {x_0}}}} 
\end{align}
and applying this functional to each side of the crossing equation. If for $\Delta>\Delta_*$, one can always find a linear functional such that 
\begin{align}
&\alpha  \circ {F_{{\Delta _0},{J_0}}}(x) > 0\nonumber\\
&\alpha  \circ {F_{\Delta ,J}}(x) > 0
\end{align}
then a contradiction is reached. This would, therefore, mean that there must exist a state with dimension $\Delta<\Delta_*$. Thus, we could search for the minimal value of $\Delta_*$, which corresponds to the gap of the theory. The search for a functional with positivity condition, can be transformed into a standard form of semidefinite programming \cite{Simmons-Duffin:2015qma}. 
\\
\\
In this paper, we will address the possibility for quantum speedup for the numerical conformal bootstrap. This is motivated for several reasons. First, some bootstrap problems are extremely difficult to solve (for instance, \cite{Dymarsky:2017yzx}) on classical computers, and it is possible that quantum computing techniques are better adapted to these questions. Second, solutions of certain numerical bootstrap problems may require higher precision in order to compare against Monte Carlo results and experiments for statistical models. Recently, rapid developments have shown that semidefinite programming indeed has the potential for dramatic quantum speedup \cite{FBe,Ae,FB,AA}. We will make use of those algorithms to address specific problems in the numerical conformal bootstrap. 
\\
\\
This paper is organized as follows. In Section \ref{revcon} we will review the conformal bootstrap, the Polynomial Matrix Program(\texttt{PMP}), and its connection to semidefinite programming(\texttt{SDP}). In Section \ref{quantum}, we will discuss the current quantum algorithms for solving \texttt{SDP}s. In Section \ref{spe} we will study the particular speedups in the context of numerical bootstrap problems by quantum algorithms. In Section \ref{ex} we will give an explicit example, the relatively well-studied three dimensional Ising bootstrap, to discuss in more specificity the potential for quantum speedup. Finally, in Section \ref{conc}, we conclude and discuss several future research directions.

\section{Conformal bootstrap and semidefinite programming}\label{revcon}
In this section, we will review some basics about the conformal bootstrap program and semidefinite programming from \cite{Simmons-Duffin:2015qma}. The conformal bootstrap problem can typically be written in the following form. Consider a given symmetric polynomial matrix set
\begin{align}
M_j^n(x) = \left( {\begin{array}{*{20}{c}}
{P_{j,11}^n(x)}&{ \cdots \;}&{P_{j,1{m_j}}^n(x)}\\
 \vdots &{ \ddots {\rm{ \;}}}&{ \vdots {\rm{ \;}}}\\
{P_{j,{m_j}1}^n(x)}&{ \cdots \;}&{P_{j,{m_j}{m_j}}^n(x)}
\end{array}} \right)
\end{align}
where $n\in \{0,1,\cdots,N\}$ and $j\in \{1,\cdots,J\}$, and the $P^n_{j,rs}(x)$ are all polynomial. Thus, given $b\in \mathbb{R}^N$, one can run the following Polynomial Matrix Program(\texttt{PMP}):
\begin{align}
\texttt{PMP: }&\text{maximize: }b\cdot y \text{ for }y\in \mathbb{R}^N\nonumber\\
&\text{such that: }M_j^0(x) + \sum\limits_{n = 1}^N {{y_n}M_j^n(x)} \succeq 0 \text{  for all }x\ge 0 \text{  and  }1\le j \le J 
\end{align}
Where the symbol $M\succeq 0$ means that $M$ is positive semi-definite.
\\
\\
Here we can give an example about how the conformal bootstrap can be transformed into the \texttt{PMP}. Consider the 3D Ising CFT. In this theory, the crossing equation for $\left\langle {\sigma \sigma \sigma \sigma } \right\rangle $, $\left\langle {\sigma \sigma \epsilon \epsilon } \right\rangle $ and $\left\langle {\epsilon \epsilon \epsilon \epsilon } \right\rangle$ will be given by
\begin{align}
\left( {\begin{array}{*{20}{c}}
1&1
\end{array}} \right){{\vec {V}}_{ + ,0,0}}\left( {\begin{array}{*{20}{c}}
1\\
1
\end{array}} \right) + \sum\limits_{{O^ + }} {\left( {\begin{array}{*{20}{c}}
{{\lambda _{\sigma \sigma O}}}&{{\lambda _{\epsilon \epsilon O}}}
\end{array}} \right)} {{\vec V}_{ + ,\Delta ,\ell }}\left( {\begin{array}{*{20}{c}}
{{\lambda _{\sigma \sigma O}}}\\
{{\lambda _{\epsilon \epsilon O}}}
\end{array}} \right) + \sum\limits_{{O^ - }} {\lambda _{\sigma \epsilon O}^2} {{\vec V}_{ - ,\Delta ,\ell }} = 0
\end{align}
Note here that $O^+$ runs over the $\mathbb{Z}_2$ even operators with even spin, while $O^-$ runs over the $\mathbb{Z}_2$ odd operators with all possible spins. Further, $\Delta$, $\ell$ are the dimension and spin of the operator $O$. The vectors $\vec V$ are then defined with the following form
\begin{align}
{{\vec V}_{ - ,\Delta ,\ell }} = \left( {\begin{array}{*{20}{c}}
0\\
0\\
{F_{ - ,\Delta ,\ell }^{\sigma \epsilon ,\sigma \epsilon }(u,v)}\\
{{{( - 1)}^\ell }F_{ - ,\Delta ,\ell }^{\epsilon \sigma ,\sigma \epsilon }(u,v)}\\
{ - {{( - 1)}^\ell }F_{ + ,\Delta ,\ell }^{\epsilon \sigma ,\sigma \epsilon }(u,v)}
\end{array}} \right)~~~~~~ {{\vec V}_{ + ,\Delta ,\ell }} = \left( {\begin{array}{*{20}{c}}
{\left( {\begin{array}{*{20}{c}}
{F_{ - ,\Delta ,\ell }^{\sigma \sigma ,\sigma \sigma }(u,v)}&0\\
0&0
\end{array}} \right)}\\
{\left( {\begin{array}{*{20}{c}}
0&0\\
0&{F_{ - ,\Delta ,\ell }^{\epsilon \epsilon ,\epsilon \epsilon }(u,v)}
\end{array}} \right)}\\
{\left( {\begin{array}{*{20}{c}}
0&0\\
0&0
\end{array}} \right)}\\
{\left( {\begin{array}{*{20}{c}}
0&{\frac{1}{2}F_{ - ,\Delta ,\ell }^{\sigma \sigma ,\epsilon \epsilon }(u,v)}\\
{\frac{1}{2}F_{ - ,\Delta ,\ell }^{\sigma \sigma ,\epsilon \epsilon }(u,v)}&0
\end{array}} \right)}\\
{\left( {\begin{array}{*{20}{c}}
0&{\frac{1}{2}F_{ + ,\Delta ,\ell }^{\sigma \sigma ,\epsilon \epsilon }(u,v)}\\
{\frac{1}{2}F_{ + ,\Delta ,\ell }^{\sigma \sigma ,\epsilon \epsilon }(u,v)}&0
\end{array}} \right)}
\end{array}} \right)\qquad 
\end{align}
while the function $F$ is given by a combination of $g$, the conformal blocks, which are known functions:
\begin{align}
F_{ \mp ,\Delta ,J}^{ij,kl}(u,v) = {v^{({\Delta _k} + {\Delta _j})/2}}g_{\Delta ,J}^{{\Delta _{ij}},{\Delta _{kl}}}(u,v) \mp {u^{({\Delta _k} + {\Delta _j})/2}}g_{\Delta ,J}^{{\Delta _{ij}},{\Delta _{kl}}}(v,u)
\end{align}
where we denote $\Delta_{ij}=\Delta_i-\Delta_j$. Now consider a functional vector $\vec{\alpha}=(\alpha^i)$, for $i=\{1,2,\cdots,5\}$. We now act with $\alpha$ on the crossing equation
\begin{align}
\left( {\begin{array}{*{20}{c}}
1&1
\end{array}} \right)\vec \alpha  \cdot {{\vec V}_{ + ,0,0}}\left( {\begin{array}{*{20}{c}}
1\\
1
\end{array}} \right) + \sum\limits_{{O^ + }} {\left( {\begin{array}{*{20}{c}}
{{\lambda _{\sigma \sigma O}}}&{{\lambda _{\epsilon \epsilon O}}}
\end{array}} \right)} \vec \alpha  \cdot {{\vec V}_{ + ,\Delta ,\ell }}\left( {\begin{array}{*{20}{c}}
{{\lambda _{\sigma \sigma O}}}\\
{{\lambda _{\epsilon \epsilon O}}}
\end{array}} \right) + \sum\limits_{{O^ - }} {\lambda _{\sigma \epsilon O}^2} \vec \alpha  \cdot {{\vec V}_{ - ,\Delta ,\ell }} = 0
\end{align}
Then, if we could find an $\alpha$ such that 
\begin{align}
&\left( {\begin{array}{*{20}{c}}
1&1
\end{array}} \right)\vec \alpha  \cdot {{\vec V}_{ + ,0,0}}\left( {\begin{array}{*{20}{c}}
1\\
1
\end{array}} \right) > 0\nonumber\\
&\vec \alpha  \cdot {{\vec V}_{ + ,\Delta ,\ell }} \ge 0\nonumber\\
&\vec \alpha  \cdot {{\vec V}_{ - ,\Delta ,\ell }} \ge 0
\end{align}
then the theory is ruled out; this method can therefore be used to rule out some range of $\Delta$'s from being in the allowed theory space.
\\
\\
In order to transform this problem to a \texttt{PMP}, we will consider the following linear functional
\begin{align}
{\alpha ^i}(f) = \sum\limits_{m \ge n,m + n \le \Lambda } {a_{mn}^i\partial _z^m\partial _{\bar z}^n} f(z,\bar z){|_{z,\bar z = 1/2}}
\end{align}
where $u=z\bar{z}$ and $v=(1-z)(1-\bar{z})$. Note that derivatives of function $F$ are well approximated by polynomials, with positive prefactors
\begin{align}
\partial _z^m\partial _{\bar z}^nF_{ \pm \Delta ,\ell }^{ij,kl}(z,\bar z){|_{z,\bar z = 1/2}} \approx {\chi _\ell }(\Delta )P_{ \pm ,\ell }^{ij,kl;mn}(\Delta )
\end{align}
where $P$s are polynomials and $\chi$s are positive prefactors. Thus we arrive at a standard form of the \texttt{PMP},
\begin{align}
\texttt{3DIsing: }&\text{find: }a_{mn}^i \nonumber\\
&\text{such that: }
\left( {\begin{array}{*{20}{c}}
1&1
\end{array}} \right){Z_0}\left( {\begin{array}{*{20}{c}}
1\\
1
\end{array}} \right) > 0\nonumber\\
&{Z_\ell }(\Delta ) \ge 0\text{ for all }\mathbb{Z}_2\text{-even operators with even spin}\nonumber\\
&{Y_\ell }(\Delta ) \ge 0\text{ for all }\mathbb{Z}_2\text{-odd operators}
\end{align}
where
\begin{align}
&{Y_\ell }(\Delta ) \equiv \sum\limits_{mn} {\left[ {a_{mn}^3P_{ - ,\ell }^{\sigma \epsilon ,\sigma \epsilon ;mn}(\Delta ) + a_{mn}^4{{( - 1)}^\ell }P_{ - ,\ell }^{\epsilon \sigma ,\sigma \epsilon ;mn}(\Delta ) - a_{mn}^5{{( - 1)}^\ell }P_{ + ,\ell }^{\epsilon \sigma ,\sigma \epsilon ;mn}(\Delta )} \right]} \nonumber\\
&{Z_\ell }(\Delta ) \equiv \sum\limits_{mn} {\left( {\begin{array}{*{20}{c}}
{a_{mn}^1P_{ - ,\ell }^{\sigma \sigma ,\sigma \sigma ;mn}(\Delta )}&{\frac{1}{2}\left( {a_{mn}^4P_{ - ,\ell }^{\sigma \sigma ,\epsilon \epsilon ;mn}(\Delta ) + a_{mn}^5P_{ + ,\ell }^{\sigma \sigma ,\epsilon \epsilon ;mn}(\Delta )} \right)}\\
{\frac{1}{2}\left( {a_{mn}^4P_{ - ,\ell }^{\sigma \sigma ,\epsilon \epsilon ;mn}(\Delta ) + a_{mn}^5P_{ + ,\ell }^{\sigma \sigma ,\epsilon \epsilon ;mn}(\Delta )} \right)}&{a_{mn}^2P_{ - ,\ell }^{\epsilon \epsilon ,\epsilon \epsilon ;mn}(\Delta )}
\end{array}} \right)} 
\end{align}
Note that $\Delta$ could be shifted to some minimal value $\Delta_\text{min}(\ell)$ plus $x$, and that we can demand $x\ge 0$ to replace the requirement that $\Delta\ge \Delta_\text{min}(\ell)$.
\\
\\
There are several things to note in this example. First, we only care about feasibility, and therefore we are allowed to set the objective function to zero in \texttt{PMP}. Other bootstrap problems, for instance the finding of bounds over OPE coefficients, could in general need a nontrivial objective function. Second, the set of spin is infinitely large, and so we should truncate it to be a finite set, for instance ranging from 0 towards $\ell_\text{max}$. For the bootstrap, if $\ell_\text{max}$ is sufficiently large, the positivity for all spins is typically ensured.
\\
\\
Finally, we will review how to relate \texttt{PMP}'s to \texttt{SDP}'s. Here we will use \texttt{SDP}'s of the following form:
\begin{align}
\texttt{SDP: }&\text{maximize: }\text{Tr}(CY)+b \cdot y \text{ on }y\in \mathbb{R}^N,~Y\in\mathcal{S}^K\nonumber\\
&\text{such that: }\text{Tr}(A_*Y)+By=c\nonumber\\
&\text{and }Y\succeq 0
\end{align}
for given $c\in \mathbb{R}^P$, $B\in \mathbb{R}^{P\times N}$, and $A_1, A_2,\cdots,A_P, C\in \mathcal{S}^K$. Here $S^K$ is the space of the real $K\times K$ symmetric matrix, and we write $A_*=(A_1, A_2, \cdots A_P)$. To transform our \texttt{PMP} problem into this form, we will eventually set $C=0$, although we will allow for it to remain general for now.
\\
\\
We quote the theorem for transforming $\texttt{{PMP}}$ to \texttt{SDP} reviewed and discussed in \cite{Simmons-Duffin:2015qma}:
\begin{theorem}
Given a \texttt{PMP} problem, if one sets $p$ to be tuples $(i,j,k,l)$ with $0\le r \le s<m_j$, $0\le k \le d_j$ and $1\le j\le J$, then this tranforms to an \texttt{SDP} of the form 
\begin{align}
&{A_{(j,r,s,k)}} = \left( {\begin{array}{*{20}{c}}
0& \cdots &0&0& \cdots &0\\
 \vdots & \ddots & \vdots & \vdots &{}& \vdots \\
0& \cdots &{{Q_{{\delta _{j1}}}}({x_k}) \otimes {E^{rs}}}&0& \cdots &0\\
0& \cdots &0&{{x_k}{Q_{{\delta _{j2}}}}({x_k}) \otimes {E^{rs}}}& \cdots &0\\
 \vdots &{}& \vdots & \vdots & \ddots & \vdots \\
0& \cdots &0&0& \cdots &0
\end{array}} \right)\nonumber\\
&{B_{(j,r,s,k),n}} =  - P_{j,rs}^n({x_k})\nonumber\\
&{c_{(j,r,s,k)}} = P_{j,rs}^0({x_k})\nonumber\\
&C = 0
\end{align}
where $d_j=\max_{n=0}^N(\deg(M_j^n(x)))$, $\delta_{j1}=\left\lfloor {{d_j}/2} \right\rfloor $ and $\delta_{j2}=\left\lfloor {{(d_j-1)}/2} \right\rfloor$ and the $x_k$ are some chosen points. We also define $(E^{rs})_{ij}=\frac{1}{2}(\delta^r_i\delta^s_j+\delta^s_i\delta^r_j)$ and $Q_\delta(x)=\vec{q}_\delta(x)\vec{q}_\delta(x)^T$, with $\vec{q}_\delta=(q_0(x),\cdots q_\delta (x))$, is a collection of polynomials with degree $0$, $1,\cdots, \delta$.
\end{theorem}
Based on the theorem, we know that the volume of matrices, as input, are given by
\begin{align}
&P=\mathcal{O}(Jm^2d)\nonumber\\
&K=\mathcal{O}(Jmd)
\end{align}
where we assume that $M$ is an $m\times m$ matrix with maximal degree $d$, and $N$ in the \texttt{PMP} is the same as that of the \texttt{SDP}.
\\
\\
There are several polynomial time classical algorithms for solving \texttt{SDP} problems. In the current conformal bootstrap community, people often use \texttt{SDPB}, an efficient \texttt{PMP} solver proposed in \cite{Simmons-Duffin:2015qma} to solve bootstrap problems based on the above transformation from \texttt{PMP} to \texttt{SDP}. The \texttt{SDP} solver in \cite{Simmons-Duffin:2015qma} is based on the exact interior point method. The complexity of the algorithm, within a reasonably small error, is known to be $\mathcal{O}(Jm^6d^3+N^2Jm^2d+N^3)$.
\section{Quantum algorithms}\label{quantum}
Recently rapid progress has been made in finding a quantum version of an \texttt{SDP} solver. The first quantum \texttt{SDP} solver was given in \cite{FBe}, and further developments refinements and speedups were found in \cite{FB}. In addition, an independent group obtained related results with different formulations and input models \cite{Ae,AA}. In this section, we will describe two quantum algorithms most following \cite{FB} for solving \texttt{SDP} that could be useful for the numerical conformal bootstrap, while an improvement with a stronger input model in \cite{AA} is also mentioned. 
\\
\\
When constructing quantum algorithms, we first need to specify how to input the elements of the matrices and vectors for a given $\texttt{SDP}$ into the quantum computer, namely, how to compile the information of matrix and vectors into the quantum circuits. Those are called \emph{Input models} or \emph{Oracles}. The first algorithm we will describe uses an oracle called the plain model, while in the second algorithm uses a quantum input model. Those two oracles, perhaps unsurprisingly, lead to algorithms with different complexities. The quantum input model is a stronger oracle that applies in a more restricted set of cases, and thus the corresponding algorithm is faster in its scaling with certain parameters. If we have a good quantum computer, which oracle one chooses will sensitively depend on the detailed quantum architecture at hand.
\\
\\
Before going into detailed discussion, let us first remark on two issues. First, in the current literature for quantum \texttt{SDP} solvers, people use the language of complex(Hermitian) matrices instead of real(symmetric) matrices. However, with some simple modifications (for instance, replacing the Hermitian matrix with a real symmetric matrix in the proof of theorems, or using a real Gibbs sampler instead of a complex one, etc.), one can easily adapt those algorithms without changing the complexity analysis. 
\\
\\
Second, people use different forms to describe equivalent \texttt{SDP} problems. In the following descriptions, the first two algorithms are written in terms of \texttt{SDP} with the form of the Semidefinite Programming (Approximate) Feasibility Problem (\texttt{SDPFP})
\begin{align}
\texttt{SDPFP: }&\text{Find }\text{Tr}(\hat{A}_{\hat{i}}\hat{X})\le \hat{a}_{\hat{i}}+\epsilon\text{ for all }\hat{i}\in \{1,2,\cdots,\hat{m}\}\nonumber\\
&\text{such that }\hat{X}\succeq 0~~~~~~\text{Tr}(\hat{X})=1\nonumber\\
&\text{for given }\epsilon>0~,~\hat{A}_{\hat{i}}\in\mathcal{S}^{\hat{n}}~,~-I\preceq \hat{A}_{\hat{i}}\preceq I
\end{align}
Here, we use the hat notation to distinguish from the variables we used above. The above conditions define a convex region $S_\epsilon$ for given $\epsilon$. The result of the algorithm will be: (1) If $S_0=\varnothing $ we will output \texttt{Fail}; else, (2) if $S_0\ne \varnothing $, output an $\hat{X}$. In the above notation, $\hat{m}$ is the number of constraints, $\hat{n}$ is the number of dimensions, and $\epsilon$ denotes the additive error of a solution. Further, we use $A\succeq B$ to denote the condition that $A-B$ is  positive semidefinite, while similarly $A\preceq B$ denotes that $B-A$ is positive semidefinite. $I$ is the identity matrix. We choose the dimension of matrix $\hat{n}$, the number of constraints $\hat{m}$, and the number \emph{sparsity} (namely, take the maximal number of non-zero elements in each row of each matrix) of $\hat{A}_i$ to be $\hat{s}$.  
\\
\\
Several versions of \texttt{SDP} problems are equivalent and simply related. We discuss them in Appendix \ref{A}.
\subsection{Quantum \texttt{SDP} solver with plain model}
Here we describe our first algorithm, the quantum solver for \texttt{SDPFP} in \cite{FB}. This algorithm is based on the following building blocks,
\\
\\
\textbf{Oracle} (plain model): The plain model is a quantum oracle $\mathcal{P}_{\hat{A}}$ that gives the map
\begin{align}
\ket{\hat{j},\hat{k},\hat{l},\hat{z}}\to \ket{\hat{j},\hat{k},\hat{l},\hat{z} \oplus (\hat{A}_{\hat{j}}) _{\hat{k}\hat{f}_{\hat{j}\hat{k}}(\hat{l})}}
\end{align}
where $\hat{j} \in \{1,2,\cdots,\hat{m}\}$, $\hat{k} \in \{1,2,\cdots,\hat{n}\}$ and $\hat{l} \in \{1,2,\cdots,\hat{s}\}$. Here $\hat{z}$ is a auxiliary bit, and $\hat{A}$ is the bit string representation. $\hat{f}_{\hat{j}\hat{k}}$ is a function with indices $\hat{j}$ and $\hat{k}$ that gives the $\hat{l}$th non-zero element in the row $\hat{k}$.
\\
\\
\textbf{Trace estimator}: The input of the trace estimator is a real symmetric matrix $H$, satisfying $||H||\le\Gamma$, and a density matrix $\rho$. Then one can compute $\text{Tr}(H\rho)$, with additive error $\epsilon$, and success rate at least $2/3$. We write $\mathcal{S}_\text{Tr}(\hat{s},\Gamma,\epsilon)$ to denote the sample complexity.
\\
\\
\textbf{Gibbs sampler}: The Gibbs sampler takes as input a real symmetric matrix $H$, satisfying $||H||\le\Gamma$ and prepares as output a Gibbs state $\rho=\frac{e^{-H}}{\text{Tr}(e^{-H})}$ with additive error $\epsilon$ with the help of plain model oracle. We call this Gibbs sampler $\text{Gibbs}_{\text{plain}}(\hat{s},\Gamma,\epsilon)$.
\\
\\
Now based on those building blocks, we will briefly describe the algorithm constructed in \cite{FB}.
\begin{algorithm}\label{plain}
Quantum \texttt{SDP} solver with plain model:
\begin{itemize}
\item Choose an initial weight matrix $W^{(1)}=I_{\hat{n}}$, and choose $T=\frac{16\hat{n}}{\epsilon^2}$.
\item Start a loop, that for $t=1$, $t\le T$, $t=t+1$:
\begin{itemize}
\item Apply the Gibbs sampler $\mathcal{O}(\log \hat{m}\times \mathcal{S}_{\operatorname{Tr}}(\frac{\hat{s}\log{\hat{n}}}{\epsilon^2},\frac{\log\hat{n}}{\epsilon},\epsilon))$ times, and get several $\rho^{(t)}=\frac{W^{(t)}}{\operatorname{Tr}(W^{(t)})}$. We use the sampler $\operatorname{Gibbs}_{\operatorname{plain}}{\left(\frac{\hat{s} \log \hat{n}}{\epsilon^{2}}, \frac{\log \hat{n}}{\epsilon}, \epsilon\right)}$.
\item Use these copies of $\rho^{(t)}$ to find a $\hat{j}^{(t)}$ such that $\operatorname{Tr}(\hat{A}_{\hat{j}^{(t)}}\rho^{(t)})>\hat{a}_{\hat{j}^{(t)}}+\epsilon$, where we use the trace estimator to compute the trace. If we find such $\hat{j}^{(t)}$, we update $M^{(t)}=\frac{1}{2}(I_n-\hat{A}_{\hat{j}^{(t)}})$. If we cannot find it, then we say that $\mathcal{S}_\epsilon$ is not empty and output $\rho^{(t)}$ as $\hat{X}$, the solution.
\item We update $W^{(t+1)}=\exp(\frac{\epsilon}{4}\times\sum_{\tau=1}^{t}M^{(\tau)})$.
\end{itemize}
\item We say that $\mathcal{S}_\epsilon$ is empty.
\end{itemize}
\end{algorithm}
We note that there is an important procedure for finding a suitable $\hat{j}^{(t)}$. This procedure can be implemented by a fast quantum OR lemma. For details, see \cite{FB}. 
\\
\\
The gate complexity we need for running this algorithm is given by $\mathcal{O}(\hat{s}^2(\frac{\sqrt{\hat{m}}}{\epsilon^{10}}+\frac{\sqrt{\hat{n}}}{\epsilon^{12}}))$, if we use the real version of the Gibbs sampler from \cite{sam} and the trace estimator from \cite{FBe}.
\\
\\
In \cite{AA}, the corresponding algorithm is similar, but they use a different but stronger input model, the quantum operator model, defined by the following.
\\
\\
\textbf{Oracle} (quantum operator model): For matrices $\hat{A}_{\hat{j}}$, find an operator $U$ such that ${\hat A_{\hat j}} = \alpha ({\left\langle 0 \right|^{ \otimes a}} \otimes I){U_{\hat j}}({\left| 0 \right\rangle ^{ \otimes a}} \otimes I)$ for some given $\alpha$ and $a$. The oracle $O_U$ is then defined such that $O_U\ket{\hat{j},\psi}=\ket{\hat{j}}U_{\hat{j}}\ket{\psi}$.
\\
\\
With this input model, the runtime required is reduced to $\mathcal{O}(\hat{s}^2(\frac{\sqrt{\hat{m}}}{\epsilon^{4}}+\frac{\sqrt{\hat{n}}}{\epsilon^{5}}))$.
\subsection{Quantum \texttt{SDP} solver with quantum input model}
Now let us consider another quantum solver, this time with a quantum input model. Taking advantage of this input model, the gate complexity will be reduced for some input parameters. 
\\
\\
We assume that $\hat{A}_{\hat{j}}$ can be decomposed as $\hat{A}_{\hat{j}}=\hat{A}_{\hat{j}}^+-\hat{A}_{\hat{j}}^-$, while $\hat{A}_{\hat{j}}^\pm \succeq 0$.This assumption will always hold when $\hat{A}_{\hat{j}}$ is a real symmetric matrix. In fact, one can take $\zeta$ large enough such that $\hat{A}_{\hat{j}}+\zeta I$ is a diagonally dominant matrix. Since it is diagonally dominant and it is symmetric, it is positive semidefinite. Thn we can simply take $\hat{A}_{\hat{j}}+\zeta I=\hat{A}_{\hat{j}}^+$ and $\zeta I=\hat{A}_{\hat{j}}^-$. In our problem, we know the input is bounded by
\begin{align}
-I\preceq \hat{A}_{\hat{i}}\preceq I
\end{align}
A naive choice is to set $\zeta=1$; better decompositions may exist, however, for specific problems.
\\
\\
We will now describe the building blocks of the algorithm.
\\
\\
\textbf{Oracle} (quantum input model for trace of $\hat{A}_{\hat{j}}$): We define a quantum oracle $O_\text{Tr}$ such that $O_\text{Tr}\ket{\hat{j},0,0}=\ket{\hat{j},\hat{A}_{\hat{j}}^+,\hat{A}_{\hat{j}}^-}$ for all $\hat{j}$.
\\
\\
\textbf{Oracle} (quantum input model for $\hat{A}_{\hat{j}}$): We define a quantum oracle $O$ such that 
\begin{align}
O\left| {\hat j} \right\rangle \left\langle {\hat j} \right| \otimes \left| 0 \right\rangle \left\langle 0 \right| \otimes \left| 0 \right\rangle \left\langle 0 \right|{O^\dag } = \left| {\hat j} \right\rangle \left\langle {\hat j} \right| \otimes \left| {\psi _{\hat j}^ + } \right\rangle \left\langle {\psi _{\hat j}^ + } \right| \otimes \left| {\psi _{\hat j}^ - } \right\rangle \left\langle {\psi _{\hat j}^ - } \right|
\end{align}
where $\ket{\psi^{\pm}_{\hat{j}}}$ are purifications of $\frac{\hat{A}^\pm_{\hat{j}}}{\operatorname{Tr}{\hat{A}^\pm_{\hat{j}}}}$.
\\
\\
\textbf{Oracle} (quantum input model for $\hat{a}_{\hat{j}}$): We define a quantum oracle $O_{\hat{a}}$ such that 
\begin{align}
O_{\hat{a}}\ket{\hat{j}}\bra{\hat{j}} \otimes \ket{0}\bra{0}O_{\hat{a}}^\dagger=\ket{\hat{j}}\bra{\hat{j}}\otimes \ket{\hat{a}_{\hat{j}}}\bra{\hat{a}_{\hat{j}}}
\end{align}
\\
We assume that in the input we have $\hat{A}_{\hat{j}}$ has rank at most $\hat{r}$ and $\operatorname{Tr}(\hat{A}_{\hat{j}}^+)+\operatorname{Tr}(\hat{A}_{\hat{j}}^-)\le \hat{B}$ where $\hat{B}$ can be understood as the upper bound of the norm for all inputs. In our problem we have 
\begin{align}
\hat B \le  \mathcal{O}(\dim I) = \mathcal{O}(\hat n)
\end{align}
Moreover, we define
\\
\\
\textbf{Trace estimator}: The input of the trace estimator is a state $\rho$, and the output will give, within success probability $1-\mathcal{O}(1/\hat{m})$, whether $\operatorname{Tr}(\hat{A}_{\hat{j}}\rho)>\hat{a}_{\hat{j}}+\epsilon$ or $\operatorname{Tr}(\hat{A}_{\hat{j}}\rho)<\hat{a}_{\hat{j}}$. We denote $\mathcal{S}_{\operatorname{Tr}}(\hat{B},\epsilon)$ as the sample complexity. The estimator is defined based on the quantum input models defined above.
\\
\\
\textbf{Gibbs sampler}: The Gibbs sampler is defined as, we assume $\hat{K}=\hat{K}^++\hat{K}^-$, where $\hat{K}^\pm=\sum_{\hat{j}\in\hat{\mathcal{S}}}\hat{c}_{\hat{j}}A_{\hat{j}}^{\pm}$. Here $\hat{\mathcal{S}}$ is a subset of $\{1,2,\cdots,\hat{m}\}$, and it satisfies $\abs{\hat{\mathcal{S}}}\le \Phi$, and coefficients $\hat{c}_{\hat{j}}>0$. Then if we know $\operatorname{Tr}(\hat{K}^+)+\operatorname{Tr}(\hat{K}^-)\le \hat{B}_{\hat{K}}$, and we have that the rank of $\hat{K}^\pm$ is at most $\hat{r}_{\hat{K}}$, then the Gibbs sampler will prepare the Gibbs state $\rho_{\text{G}}=\text{exp} (-\hat{K})/\operatorname{Tr}(\text{exp} (-\hat{K}))$ with error $\epsilon$, using the quantum input oracles above. We call this Gibbs sampler $\text{Gibbs}_{\text{quantum}}\left(\hat{r}_{\hat{K}}, \Phi, \hat{B}_{\hat{K}}, \epsilon\right)$.
\\
\\ 
Based on those building blocks we have
\begin{algorithm}\label{Q}
Quantum \texttt{SDP} solver with quantum input model:
\begin{itemize}
\item Choose an initial weight matrix $W^{(1)}=I_{\hat{n}}$, and choose $T=\frac{16\hat{n}}{\epsilon^2}$.
\item Start a loop, that for $t=1$, $t\le T$, $t=t+1$:
\begin{itemize}
\item Apply the Gibbs sampler $\mathcal{O}(\mathcal{S}_{\operatorname{Tr}}(\hat{B},\epsilon))$ times, and get several $\rho^{(t)}=\frac{W^{(t)}}{\operatorname{Tr}(W^{(t)})}$. We use the sampler $\operatorname{Gibbs}_{\operatorname{quantum}}\left(\frac{\hat{r} \log \hat{n}}{\epsilon^{2}}, \frac{16 \log \hat{n}}{\epsilon^{2}}, \frac{\hat{B} \log \hat{n}}{\epsilon}, \epsilon\right)$.
\item Use these copies of $\rho^{(t)}$ to find a $\hat{j}^{(t)}$ such that $\operatorname{Tr}(\hat{A}_{\hat{j}^{(t)}}\rho^{(t)})>\hat{a}_{\hat{j}^{(t)}}+\epsilon$, where we use the trace estimator to compute the trace. If we find such a $\hat{j}^{(t)}$, we then update $M^{(t)}=\frac{1}{2}(I_n-\hat{A}_{\hat{j}^{(t)}})$. If we cannot find it, then we say that $\mathcal{S}_\epsilon$ is not empty and output $\rho^{(t)}$ as $\hat{X}$, the solution.
\item We update $W^{(t+1)}=\exp(\frac{\epsilon}{4}\times\sum_{\tau=1}^{t}M^{(\tau)})$.
\end{itemize}
\item We say that $\mathcal{S}_\epsilon$ is empty.
\end{itemize}
\end{algorithm}
This algorithm is very efficient based on the Gibbs sampler and the trace estimator that are constructed in \cite{FB}. Its complexity is given by
\begin{align}
&\mathcal{O}((\sqrt{\hat{m}} + \text{poly}(\hat{r})) \times \text{poly}(\log \hat{m},\log \hat{n},\hat{B},{\epsilon^{ - 1}}))\nonumber\\
&\le \mathcal{O}((\sqrt{\hat{m}} + \text{poly}(\hat{r})) \times \text{poly}(\log \hat{m},\log \hat{n},\mathcal{O}(\hat{n}),{\epsilon^{ - 1}}))\
\end{align}
A remarkable fact here is that it only has the polylog dependence $\hat{n}$, and square-root $\times$ polylog dependence on $\hat{m}$ regardless of $\hat{B}$.
\section{Specifying bootstrap problems}\label{spe}
When discussing the bootstrap problem, we can start from \texttt{PMP} problems. For the \texttt{PMP} parameter setup, we estimate that the \texttt{SDPFP} parameters are (For transformations of several forms of \texttt{SDP}, see Appendix \ref{A})
\begin{align}
&\hat m = \mathcal{O}(J{m^2}d)\nonumber\\
&\hat n = \mathcal{O}(Jmd + N)
\end{align}
Moreover, assuming a generic \texttt{PMP} setup (namely assuming that the inputs of derivatives are not empty), the sparsity is 
\begin{align}
\hat s=\mathcal{O}(d)
\end{align}
Classical algorithms are able to solve \texttt{PMP} and \texttt{SDP} in polynomial time. The current practical programming tool, \texttt{SDPB}, has runtime estimated to be $\mathcal{O}(Jm^6d^3+N^2Jm^2d+N^3)$. If one uses Arora and Kale's multiplicative weight method \cite{AK} to solve \texttt{PMP}, it gives $\mathcal{O}({J^2}{m^3}{d^3} + NJ{m^2}{d^2})$ if we only care about manifest dependence on numbers, dimensions and sparsities of matrices\footnote{Here those two algorithms we mention are standard examples, but one could in principle consider more types of classical algorithms.}.
\\
\\
Similarly, the quantum algorithm with plain input model works for the \texttt{SDP} problem, and thus works for \texttt{PMP} problems. The runtime estimate here would be
\begin{align}
\mathcal{O}({d^{2.5}}{J^{0.5}}m + {J^{0.5}}{m^{0.5}}{d^{2.5}} + {d^2}{N^{0.5}})
\end{align}
Thus, generically the quantum algorithm with plain input model will provide a quantum speedup.
\\
\\
For the quantum input model, we know that the rank of the input $\hat{A}_{\hat{j}}$ is at most,
\begin{align}
\hat r \le \mathcal{O}(d + N)
\end{align}
Thus in the quantum input model, the time cost will be
\begin{align}
&\mathcal{O}((m{d^{0.5}}{J^{0.5}} + {\rm{poly(}}\hat r)) \times {\rm{poly}}(\log (J),\log (m),\log (d),\log (N),\hat B))\nonumber\\
&\le \mathcal{O}((m{d^{0.5}}{J^{0.5}} + {\rm{poly(}}d + N{\rm{)}}) \times {\rm{poly}}(\log (J),\log (m),\log (d),\log (N),\mathcal{O}(Jmd + N)))
\end{align}
Now let us consider some specific cases:
\begin{itemize}
\item If we assume that $\hat{r}$ and $\hat{B}$ are small enough, in this case the time cost is
\begin{align}
\mathcal{O}(m{d^{0.5}}{J^{0.5}} \times {\rm{poly}}(\log (J),\log (m),\log (d),\log (N)))
\end{align}
So it is always faster than the two given classical algorithms. For large $J$, it is slower than the plain input model, but it is faster for large $d$, $m$, $N$ (holding all other parameters fixed).
\item In the worst case we consider $\hat r = \mathcal{O}(d + N)$ and $\hat{B}=\mathcal{O}(Jmd+N)$. We denote the degree of polynomial dependence on $\hat{r}$ and $\hat{B}$ to be $d_r$ and $d_B$ separately. Thus, whether the algorithm is fast or not, is based on the following conditions
\begin{center}
    \begin{tabular}{| l | l | l | l | l |}
    \hline
    Requirement & large $d$ & large $J$ & large $m$& large $N$ \\ \hline
    Faster than Arora-Kale & ${d _r} + {d _B} < 2$ & ${d _B} < 1.5$ & ${d _B} < 2$& ${d _r} + {d _B} < 1$   \\ \hline
    Faster than \texttt{SDPB} & ${d _r} + {d _B} < 2$ & ${d _B} < 0.5$ & ${d _B} < 5$ & ${d _r} + {d _B} < 3$ \\ \hline
    Faster than plain model & ${d _r} + {d _B} < 1.5$ &\text{not possible} &\text{not possible}& ${d _r} + {d _B} < 0.5$ \\
    \hline
    \end{tabular}
\end{center} 
\end{itemize}
Now we take a look at the bootstrap problem itself. A typical bootstrap problem\footnote{The actual parameters that generically work for Ising and many other models are given by the appendix of \cite{Simmons-Duffin:2015qma}.} will have the parameter set
\begin{align}
&N = \mathcal{O}({\Lambda ^2})\nonumber\\
&d = \mathcal{O}({\Lambda})\nonumber\\
&J = \mathcal{O}(J)\nonumber\\
&m =\mathcal{O}(1)
\end{align}
where $\Lambda$ is the cutoff of the number of derivatives. It scales as a square because we have both $z$ and $\bar{z}$. Thus we have complexities
\begin{align}
\texttt{SDPB}\text{: }&\mathcal{O}(J\Lambda^3+\Lambda^6)\nonumber\\
\text{classical Arora-Kale: }&\mathcal{O}(J^2\Lambda^3+J\Lambda^4)\nonumber\\
\text{plain input model: }&\mathcal{O}(J^{0.5}\Lambda^{2.5}+\Lambda^3)
\end{align}
And generically we expect
\begin{align}
\text{quantum input model: }&{\mathcal{ O}}((\Lambda {J^{0.5}} + {\rm{pol}}{{\rm{y}}_r}(\mathcal{O}({\Lambda ^2}))){\rm{poly}}(\log J,\log \Lambda ,\mathcal{O}(J{\Lambda }+\Lambda^2)))
\end{align}
Thus for the plain input model, generically we have speedup for all parameters, while for the quantum input model we have, in the worst case,
\begin{center}
    \begin{tabular}{ | l | l | l | }
    \hline
    Requirement & large $\Lambda$ & large $J$  \\ \hline
    Faster than Arora-Kale & ${d _B} < 1.5$ and $d_r+d_B<2$ & ${d _B} < 1.5$  \\ \hline
    Faster than \texttt{SDPB} & ${d _B} < 2.5$ and $d_r+d_B<3$ & ${d _B} < 0.5$\\ \hline
    Faster than plain model & ${d _B} < 1$ and $d_r+d_B<1.5$&\text{not possible}\\
    \hline
    \end{tabular}
\end{center} 
We should comment briefly on those results. First, we see that requiring a quantum speed up and advantage over the plain input model strongly constrains the allowed quantum oracle algorithms. The reason is that for bootstrap problems, we typically require the input matrix to have a large rank, which effectively negates the exponential advantage of the quantum oracle algorithm. On the other hand, bootstrap problems often have low $\hat{s}$ (namely, the input is very sparse), because the sparsity $\hat{s}$ has a very weak dependence on the large numbers (linear in $d$). That means that the square root speedup of the plain input model algorithm will be generically robust for such problems.

\section{Examples}\label{ex}
\begin{figure}[htbp]
  \centering
  \includegraphics[width=0.8\textwidth]{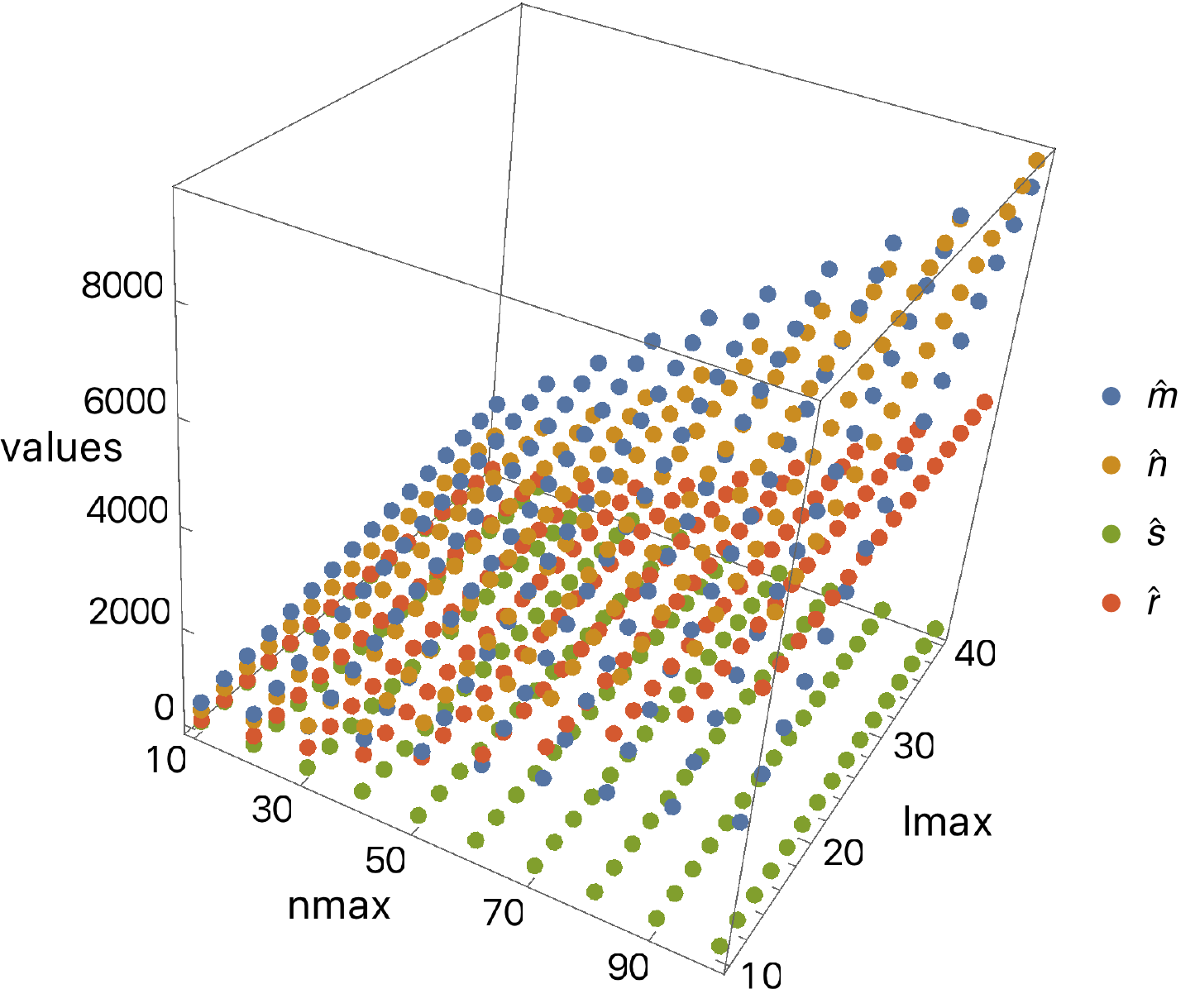}
  \caption{\label{fig1} Analytic expression of parameters in \texttt{SDPFP} for the \texttt{3DIsing(reduced)} setup.}
\end{figure}

\begin{figure}[htbp]
  \centering
  \includegraphics[width=0.8\textwidth]{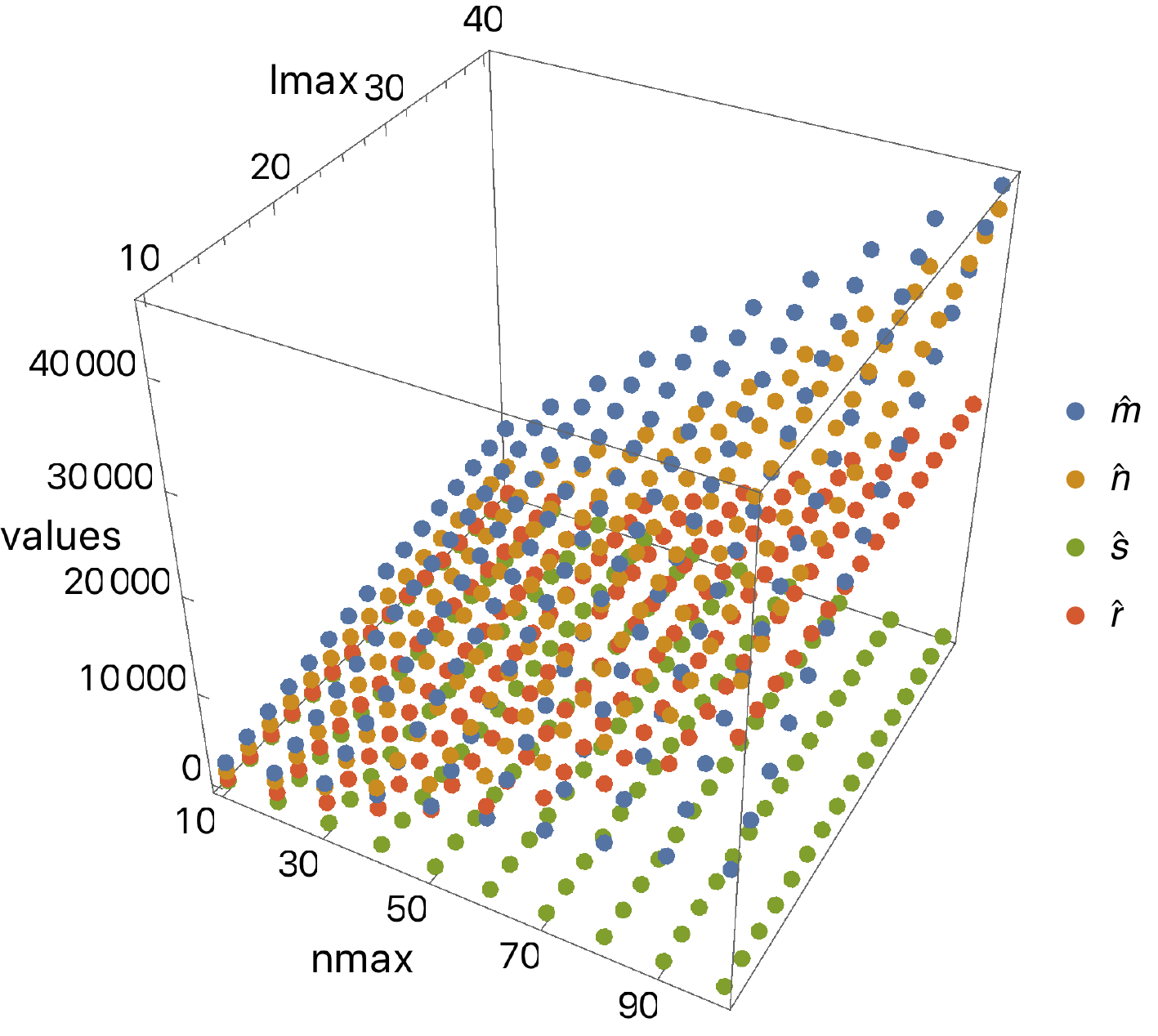}
  \caption{\label{fig2} Analytic expression of parameters in \texttt{SDPFP} for the \texttt{3DIsing} setup.}
\end{figure}

It is valuable to know how parameters scale in the actual bootstrap problems. We will provide some examples here.
\\
\\
\textbf{Crossing of $\left<\sigma \sigma \sigma \sigma\right>$ in 3D Ising: }Here we consider only the first crossing equation in the \texttt{3DIsing} problem. In this special case, positivity of matrices become positivity of numbers. Namely, the problem now becomes 
\begin{align}
\texttt{3DIsing(reduced): }&\text{find: }\tilde{a}_{mn} \nonumber\\
&\text{such that }\tilde{Z}_0> 0\nonumber\\
&{\tilde{Z}_\ell }(\Delta ) >0\text{ for all }\mathbb{Z}_2\text{-even operators with even spin}
\end{align}
where
\begin{align}
{\tilde{Z}_\ell }(\Delta ) \equiv \sum\limits_{mn}\tilde{a}_{mn}P_{ - ,\ell }^{\sigma \sigma ,\sigma \sigma ;mn}(\Delta )
\end{align}
In this problem, we will label the maximal number of derivatives to be \texttt{nmax}, while the maximal spin we take to be \texttt{lmax} (The relation between \texttt{nmax} and $\Lambda$ is given by $\Lambda=2\times\texttt{nmax}-1$). For simplicity, we consider \texttt{lmax} to be even and take the parameters to not be too small. The parameters in the theory will then scale as the following,
\begin{center}
    \begin{tabular}{ | l | l |  }
    \hline
    Parameters & Values   \\ \hline
     $J$ & $0.5\times \texttt{lmax}+1$ \\  \hline
    $m$ & $1$ \\  \hline
    $d$ & $2\times \texttt{nmax}+18$ \\  \hline
    $N$ & $0.5\times \texttt{nmax}^2+0.5\times \texttt{nmax}+1$  \\  \hline
    $P$ & $\texttt{nmax}\times\texttt{lmax}+9.5\times \texttt{lmax}+2\times\texttt{nmax}$  \\ \hline
    $K$& $\texttt{nmax}\times\texttt{lmax}+9.5\times \texttt{lmax}+2\times\texttt{nmax}$ \\ \hline
    $\hat{m}$ & $2\times \texttt{nmax}\times\texttt{lmax}+19\times \texttt{lmax}+4\times\texttt{nmax}+1$ \\  \hline
    $\hat{n}$ & $0.5\times \texttt{nmax}^2+\texttt{nmax}\times\texttt{lmax}+9.5\times \texttt{lmax}+2.5\times\texttt{nmax}+1$ \\  \hline
    $\hat{s}$ & $\texttt{nmax}+10$ \\  \hline
    $\hat{r}$ & $0.5\times \texttt{nmax}^2+0.5\times \texttt{nmax}+2$ \\  \hline
    \end{tabular}
\end{center} 
The results agree with out expectation of the generic analysis in Section \ref{spe}. We plot the \texttt{SDPFP} parameter $\hat{m},\hat{n},\hat{s},\hat{r}$ in Figure \ref{fig1}.
\\
\\
\textbf{The full 3D Ising problem: } A more serious analysis can be done for the full \texttt{3DIsing} problem. Note that here in the crossing equation we have to consider the possibility that $O^-$ is equal to $\sigma$ and $O^+$ is equal to $\epsilon$ in the sum. As a result we obtain a similar table,
\begin{center}
    \begin{tabular}{ | l | l |  }
    \hline
    Parameters & Values   \\ \hline
     $J$ & $1.5\times \texttt{lmax}+3$ \\  \hline
    $m$ & $2$ \\  \hline
    $d$ & $2\times \texttt{nmax}+19$ \\  \hline
    $N$ & $2.5\times \texttt{nmax}^2+2.5\times \texttt{nmax}-1$  \\  \hline
    $P$ & $5\times \texttt{nmax}\times\texttt{lmax}+48.5\times \texttt{lmax}+8\times\texttt{nmax}-13$  \\ \hline
    $K$& $4\times \texttt{nmax}\times\texttt{lmax}+39\times \texttt{lmax}+6\times\texttt{nmax}-14$ \\ \hline
    $\hat{m}$ & $10\times \texttt{nmax}\times\texttt{lmax}+97\times \texttt{lmax}+16\times\texttt{nmax}-25$ \\  \hline
    $\hat{n}$ & $2.5\times \texttt{nmax}^2+4\times \texttt{nmax}\times\texttt{lmax}+39\times \texttt{lmax}+8.5\times\texttt{nmax}-13$ \\  \hline
    $\hat{s}$ & $\texttt{nmax}+10$ \\  \hline
    $\hat{r}$ & $\approx 2.5\times \texttt{nmax}^2+2.5\times \texttt{nmax}+2$ \\  \hline
    \end{tabular}
\end{center} 
Similarly, we plot the $\hat{m},\hat{n},\hat{s},\hat{r}$ parameter in Figure \ref{fig2}. The value of $N$ matches the dimension of $\alpha$ coefficients around equation (3.15) of \cite{Simmons-Duffin:2015qma}. Moreover, we could use the parameter setup used in Figure 1 of \cite{Simmons-Duffin:2015qma}, which is given in the following table. As we mentioned before, we see that the sparsity growth is relatively slow (linear in \texttt{nmax} with low slope). This helps to retain the advantage of the plain input model, since it depends on sparsity by $\hat{s}^2$, but it depends on $\hat{m}$ and $\hat{n}$ with square roots.
\begin{center}
    \begin{tabular}{ | l | l |l| l | l |  }
    \hline
    \texttt{nmax} &10 & 14 & 18 & 22    \\ \hline
     \texttt{lmax} &28 & 38 & 56& 76 \\  \hline
     $\hat{m}$ &$6\times 10^3$ & $9\times 10^3$ & $1\times 10^4$  & $2\times 10^4$ \\  \hline
     $\hat{n}$ &$3\times 10^3$ & $4\times 10^3$ & $7\times 10^3$ &$1\times 10^4$ \\  \hline
     $\hat{s}$ &20 & 24  & 28 &32   \\  \hline
     $\hat{r}$ & $3\times 10^2$ & $5\times 10^2$ &$9\times 10^2$ & $1\times 10^3$\\  \hline
    \end{tabular}
\end{center} 
\section{Conclusion and outlook}\label{conc}
In this paper, we address the possibility of using quantum algorithms for \texttt{SDP} problems to numerically solve crossing equations for the conformal bootstrap. Generically, we show that quantum algorithms will provide significant speedup over classical \texttt{SDP} algorithms for specific conformal bootstrap setups. We expect that these methods may have possible applications in the coming NISQ era, and beyond, to help the bootstrap community obtain more accurate and faster numerical implementations. 
\\
\\
The area of quantum \texttt{SDP} is still young. In this paper, we mainly discuss two quantum algorithms. However, in the near future, it is reasonable to believe that more efficient algorithms and more realistic constructions with corresponding quantum architectures will be developed. 
\\
\\
We will leave the following possible directions to future research.
\begin{itemize}
\item Oracles. It will be important to construct oracles (unitaries) to supply the bootstrap inputs to the quantum computers, and the complexities of implementing these oracles will also constraint the overall algorithmic complexity at hand.
\item Upgrading the Gibbs sampler and the trace estimator. One could potentially develop better methods for speeding up the Gibbs sampling and trace estimation or think about how to realize them in actual quantum devices.
\item More specified solver. The bootstrap problems take a special shape of \texttt{PMP}, which is a special category of \texttt{SDP}. It is therefore hypothetically possible to construct algorithms that take advantage of this specific form to further speed up the computation.
\item Realizations. It might be useful to perform the above algorithms in some quantum simulators, or real quantum devices in the future, to empirically test whether the theoretical predictions for runtimes are perhaps too conservative. An attempt to adapt classical \texttt{SDP} solvers for the conformal bootstrap is already underway in \cite{BKLS}\footnote{It is worth mentioning that the current \texttt{SDP} solver \cite{Simmons-Duffin:2015qma} already includes several modifications to traditional \texttt{SDP} problems. One might look for future progress to upgrade it.}.

\end{itemize}

\section*{Acknowledgments}
We thank Fernando Brand$\tilde{\text{a}}$o, Richard Kueng, John Preskill and David Simmons-Duffin for discussions. NB is supported by the National Science Foundation under grant number 82248-13067-44-PHPXH and by the Department of Energy under grant number DE-SC0019380. JL is supported in part by the Institute for Quantum Information and Matter (IQIM), an NSF Physics Frontiers Center (NSF Grant PHY-1125565) with support from the Gordon and Betty Moore Foundation (GBMF-2644), by the Walter Burke Institute for Theoretical Physics, and by Sandia Quantum Optimization \& Learning \& Simulation, DOE Award \#DE-NA0003525. We thank the workshop, \emph{Next steps in Quantum Science for HEP} in Fermilab, for hosting, where parts of work were finished. 

\appendix 
\section{Transforming different SDP problems}\label{A}
This is a short note on different equivalent forms of \texttt{SDP1}(the form of \texttt{SDP} in the main text). From now on we write 
\begin{align}
\texttt{SDP1: }&\text{maximize: }\text{Tr}(C^{(1)}Y^{(1)})+b^{(1)} \cdot y^{(1)} \text{ on }y^{(1)}\in \mathbb{R}^{N^{(1)}},~Y^{(1)}\in\mathcal{S}^{K^{(1)}}\nonumber\\
&\text{such that: }{\operatorname{Tr}}(A_*^{(1)}{Y^{(1)}}) + {B^{(1)}}{y^{(1)}} = {c^{(1)}}\nonumber\\
&\text{and }Y^{(1)}\succeq 0
\end{align}
where $A^{(1)}_*=(A^{(1)}_1,\cdots,A^{(1)}_{P^{(1)}})$. Firstly we show that it could be transformed as
\begin{align}
\texttt{SDP2: }&\text{maximize: }\text{Tr}(C^{(2)}Y^{(2)})\text{ on }Y^{(2)}\in\mathcal{S}^{K^{(2)}}\nonumber\\
&\text{such that: }{\operatorname{Tr}}(A_*^{(2)}{Y^{(2)}}) = {c^{(2)}}\nonumber\\
&\text{and }Y^{(2)}\succeq 0
\end{align}
In fact we write $y^{(1)}_i=z^{(1)}_i-s^{(1)}$, where $z^{(1)}$ and $s^{(1)}$ are positive. Then we have
\begin{align}
&{\rm{Tr}}({C^{(1)}}{Y^{(1)}}) + {b^{(1)}} \cdot {y^{(1)}}\nonumber\\
&= {\rm{Tr}}({C^{(1)}}{Y^{(1)}}) + {b^{(1)}} \cdot {z^{(1)}} - {s^{(1)}}\sum\limits_i {b_i^{(1)}} \nonumber\\
&= {\rm{Tr}}\left( {\left( {\begin{array}{*{20}{c}}
{{C^{(1)}}}&{}&{}\\
{}&{{\rm{diag}}(b_i^{(1)})}&{}\\
{}&{}&{ - \sum\limits_i {b_i^{(1)}} }
\end{array}} \right)\left( {\begin{array}{*{20}{c}}
{{Y^{(1)}}}&?&?\\
?&{{\rm{diag}}(z_i^{(1)})}&?\\
?&?&{{s^{(1)}}}
\end{array}} \right)} \right)
\end{align}
and
\begin{align}
&{\rm{Tr}}(A_i^{(1)}{Y^{(1)}}) + {B^{(1)}}{y^{(1)}} - c_i^{(1)}\nonumber\\
&= {\rm{Tr}}(A_i^{(1)}{Y^{(1)}}) + B_{ij}^{(1)}y_j^{(1)} - c_i^{(1)}\nonumber\\
&= {\rm{Tr}}(A_i^{(1)}{Y^{(1)}}) + B_{ij}^{(1)}z_j^{(1)} - {s^{(1)}}\sum\limits_j^{} {B_{ij}^{(1)}}  - c_i^{(1)}\nonumber\\
&= {\rm{Tr}}\left( {\left( {\begin{array}{*{20}{c}}
{A_i^{(1)}}&{}&{}\\
{}&{{\rm{diag}}(B_{ij}^{(1)})}&{}\\
{}&{}&{ - \sum\limits_j^{} {B_{ij}^{(1)}} }
\end{array}} \right)\left( {\begin{array}{*{20}{c}}
{{Y^{(1)}}}&?&?\\
?&{{\rm{diag}}(z_j^{(1)})}&?\\
?&?&{{s^{(1)}}}
\end{array}} \right)} \right) - c_i^{(1)}
\end{align}
where ? means it is not necessary to know. So we only need to set
\begin{align}
&{C^{(2)}} = \left( {\begin{array}{*{20}{c}}
{{C^{(1)}}}&{}&{}\\
{}&{{\rm{diag}}(b_i^{(1)})}&{}\\
{}&{}&{ - \sum\limits_i {b_i^{(1)}} }
\end{array}} \right)\nonumber\\
&{A^{(2)}} = \left( {\begin{array}{*{20}{c}}
{A_i^{(1)}}&{}&{}\\
{}&{{\rm{diag}}(B_{ij}^{(1)})}&{}\\
{}&{}&{ - \sum\limits_j^{} {B_{ij}^{(1)}} }
\end{array}} \right)\nonumber\\
&{c^{(2)}} = {c^{(1)}}
\end{align}
with 
\begin{align}
K^{(2)}= K^{(1)} + N^{(1)} + 1~~~~~~P^{(2)}= P^{(1)}
\end{align}
\footnote{The inverse is trivial if we are interested in equivalence, since if we have an \texttt{SDP1} solver, we could just set $b^{(1)}$ and $B^{(1)}$ to be zero.}Now one could also show that it could be converted to
\begin{align}
\texttt{SDP3: }&\text{maximize: }\text{Tr}(C^{(3)}Y^{(3)}) \text{ on }Y^{(3)}\in\mathcal{S}^{K^{(3)}}\nonumber\\
&\text{such that: }\text{Tr}(A_*^{(3)}Y^{(3)})\le c^{(3)}\nonumber\\
&\text{and }Y^{(3)}\succeq 0
\end{align}
In fact, if we have a solver for \texttt{SDP3}, let us take $A^{(3)}$ to be $(A_i^{(2)},-A_i^{(2)})$ (Note that now the number of $A^{(3)}$ is doubled), and we set $c^{(3)}$ to be $(c^{(2)},-c^{(2)})$. That means that we impose the constraint
\begin{align}
{c^{(2)}} \le {\rm{Tr}}(A_*^{(2)}{Y^{(2)}}) \le {c^{(2)}}
\end{align}
So it gives the equality.
\\
\\
These operations mean that we could directly use the solver $\texttt{SDP3}$ to solve the problem of $\texttt{SDP1}$. We need to just set
\begin{align}
&{C^{(3)}} = \left( {\begin{array}{*{20}{c}}
{{C^{(1)}}}&{}&{}\\
{}&{{\rm{diag}}(b_i^{(1)})}&{}\\
{}&{}&{ - \sum\limits_i {b_i^{(1)}} }
\end{array}} \right)\nonumber\\
&A^{(3)}= \left( {\left( {\begin{array}{*{20}{c}}
{A_i^{(1)}}&{}&{}\\
{}&{{\rm{diag}}(B_{ij}^{(1)})}&{}\\
{}&{}&{ - \sum\limits_j^{} {B_{ij}^{(1)}} }
\end{array}} \right), - \left( {\begin{array}{*{20}{c}}
{A_i^{(1)}}&{}&{}\\
{}&{{\rm{diag}}(B_{ij}^{(1)})}&{}\\
{}&{}&{ - \sum\limits_j^{} {B_{ij}^{(1)}} }
\end{array}} \right)} \right)\nonumber\\
&c^{(3)} = (c^{(1)}, - c^{(1)})
\end{align}
and we have
\begin{align}
K^{(3)}= K^{(1)} + N^{(1)} + 1~~~~~~P^{(3)}= 2P^{(1)}
\end{align}
\footnote{If we have an \texttt{SDP2} solver, we could also solve  \texttt{SDP1}. A naive way is to scan many possible $c^{(2)}$ which are smaller or equal to $c^{(3)}$. A more efficient way is to perform the slack variable trick again, where we define
\begin{align}
\begin{array}{l}
A_*^{(2)} = \left( {\begin{array}{*{20}{c}}
{A_*^{(3)}}&0\\
0&1
\end{array}} \right)~~~~~~
{C^{(2)}} = \left( {\begin{array}{*{20}{c}}
{{C^{(3)}}}&0\\
0&0
\end{array}} \right)~~~~~~
{Y^{(2)}} = \left( {\begin{array}{*{20}{c}}
{{Y^{(3)}}}&?\\
?&y
\end{array}} \right)
\end{array}
\end{align}
which solves the problem. }Finally we will show that we could solve \texttt{SDP1} using the \texttt{SDPFP} solver,
\begin{align}
\texttt{SDPFP: }&\text{Find }\text{Tr}(\hat{A}_{\hat{i}}\hat{X})\le \hat{a}_{\hat{i}}\text{ for all }\hat{i}\in \{1,2,\cdots,\hat{m}\}\nonumber\\
&\text{such that }\hat{X}\succeq 0~~~~~~\text{Tr}(\hat{X})=1\nonumber\\
&\text{for given }\hat{A}_{\hat{i}}\in\mathcal{S}^{\hat{n}}~,~-I\preceq \hat{A}_{\hat{i}}\preceq I
\end{align}
We could start from \texttt{SDP3}. We notice that for given ${A^{(3)}}$ there exists a positive number $\lambda$ such that 
\begin{align}
-\lambda I\preceq A_i^{(3)},C^{(3)}\preceq \lambda I
\end{align}
for all $i$. Then we could also just redefine ${A^{(3)}},C^{(3)}$ by $\frac{A^{(3)},C^{(3)}}{\lambda}$ and we should also redefine $c^{(3)}$ by $\frac{c^{(3)}}{\lambda}$. For optimization, we could using binary search by assuming for instance, ${\rm{Tr}}({C^{(3)}}{Y^{(3)}}) \ge {q_{\text{bin}}}$ or namely, ${\rm{Tr}}( - {C^{(3)}}{Y^{(3)}}) \le  - {q_{\text{bin}}}$ and treat this as a component of $A^{(3)}_*$. We could increase $q_\text{bin}$ until we cannot find the optimal solution.
\\
\\
Finally, for the condition $\text{Tr}(\hat{X})=1$, we could treat it as the following. Firstly, the solver for $\text{Tr}(\hat{X})=1$ means that we have a solver for $\text{Tr}(\hat{X})=\hat{\omega}$ when $\hat{\omega}$ is an arbitrary positive number, since we could redefine $ \hat{a}_{\hat{i}}$ by $ \frac{\hat{a}_{\hat{i}}}{\hat{\omega}}$. For a given \texttt{SDP3} problem, it does not matter if we add a constraint $\text{Tr}(Y^{(3)})\le \hat{\omega}$ if $\hat{\omega}$ is sufficiently large. Then we could use the slack variable trick by adding variable $\mu\ge 0$ and transforming the inequality to the equality $\text{Tr}(Y^{(3)})+\mu= \hat{\omega}$\footnote{The inverse is also a simple problem. Since we know that the trace one condition means that \texttt{SDPFP} is a merge of \texttt{SDP2} and \texttt{SDP3}, and the norm condition could be directly implemented as input.}. 
\\
\\
As a conclusion, for given \texttt{SDP1} problem, we define
\begin{align}
&{C^{(3)}} = \left( {\begin{array}{*{20}{c}}
{{C^{(1)}}}&{}&{}\\
{}&{{\rm{diag}}(b_i^{(1)})}&{}\\
{}&{}&{ - \sum\limits_i {b_i^{(1)}} }
\end{array}} \right)\nonumber\\
&A^{(3)}= \left( {\left( {\begin{array}{*{20}{c}}
{A_i^{(1)}}&{}&{}\\
{}&{{\rm{diag}}(B_{ij}^{(1)})}&{}\\
{}&{}&{ - \sum\limits_j^{} {B_{ij}^{(1)}} }
\end{array}} \right), - \left( {\begin{array}{*{20}{c}}
{A_i^{(1)}}&{}&{}\\
{}&{{\rm{diag}}(B_{ij}^{(1)})}&{}\\
{}&{}&{ - \sum\limits_j^{} {B_{ij}^{(1)}} }
\end{array}} \right)} \right)
\end{align}
and then we find a large positive $\lambda$ such that 
\begin{align}
-\lambda I\preceq A_i^{(3)},C^{(3)}\preceq \lambda I
\end{align}
for all $i$. Then we set
\begin{align}
\hat A = \left( \begin{array}{l}
\left( {\begin{array}{*{20}{c}}
{A_i^{(1)}}&{}&{}&{}\\
{}&{{\rm{diag}}(B_{ij}^{(1)})}&{}&{}\\
{}&{}&{ - \sum\limits_j^{} {B_{ij}^{(1)}} }&{}\\
{}&{}&{}&0
\end{array}} \right)/\lambda , - \left( {\begin{array}{*{20}{c}}
{A_i^{(1)}}&{}&{}&{}\\
{}&{{\rm{diag}}(B_{ij}^{(1)})}&{}&{}\\
{}&{}&{ - \sum\limits_j^{} {B_{ij}^{(1)}} }&{}\\
{}&{}&{}&0
\end{array}} \right)/\lambda ,\\
-\left( {\begin{array}{*{20}{c}}
{{C^{(1)}}}&{}&{}&{}\\
{}&{{\rm{diag}}(b_i^{(1)})}&{}&{}\\
{}&{}&{ - \sum\limits_i {b_i^{(1)}} }&{}\\
{}&{}&{}&0
\end{array}} \right)/\lambda
\end{array} \right)
\end{align}
and
\begin{align}
\hat{a} = ({c^{(1)}}/\lambda \hat{\omega} , - {c^{(1)}}/\lambda \hat{\omega} , - {q_{{\rm{bin}}}}/\lambda \hat{\omega} )
\end{align}
where $\hat{\omega}$ could be chosen as sufficiently large, for size of solutions of given \texttt{SDP1}. And we use binary search to choose $q_{\text{bin}}$. And we have
\begin{align}
&\hat m = 2{P^{(1)}} + 1 =\mathcal{ O}({P^{(1)}})\nonumber\\
&\hat n = {K^{(1)}} + {N^{(1)}} + 2 = \mathcal{O}({K^{(1)}} + {N^{(1)}})
\end{align}


\begin{thebibliography}{99}

\bibitem{Qu}
J.~Preskill, \emph{Quantum} 2, 79 (2018), [arXiv:1801.00862 [quant-ph]].
\bibitem{Feynman:1981tf} 
  R.~P.~Feynman,
  Int.\ J.\ Theor.\ Phys.\  {\bf 21}, 467 (1982).


\bibitem{Jordan:2011ne}
  S.~P.~Jordan, K.~S.~M.~Lee and J.~Preskill,
  Science {\bf 336}, 1130 (2012)
  [arXiv:1111.3633 [quant-ph]].
\bibitem{Jordan:2011ci}
  S.~P.~Jordan, K.~S.~M.~Lee and J.~Preskill,
  Quantum Information and Computation 14, 1014-1080 (2014)
  [arXiv:1112.4833 [hep-th]].
\bibitem{Jordan:2014tma}
  S.~P.~Jordan, K.~S.~M.~Lee and J.~Preskill,
  arXiv:1404.7115 [hep-th].
\bibitem{Jordan:2017lea}
  S.~P.~Jordan, H.~Krovi, K.~S.~M.~Lee and J.~Preskill,
  arXiv:1703.00454 [quant-ph].

\bibitem{JP}
J.~Liu,  J.~Preskill and B.~\c{S}ahino\u{g}lu, to appear.

\bibitem{Brown:2015bva}
  A.~R.~Brown, D.~A.~Roberts, L.~Susskind, B.~Swingle and Y.~Zhao,
  Phys.\ Rev.\ Lett.\  {\bf 116}, no. 19, 191301 (2016)
  [arXiv:1509.07876 [hep-th]].

\bibitem{Brown:2015lvg}
  A.~R.~Brown, D.~A.~Roberts, L.~Susskind, B.~Swingle and Y.~Zhao,
  Phys.\ Rev.\ D {\bf 93}, no. 8, 086006 (2016)
  [arXiv:1512.04993 [hep-th]].

\bibitem{Brown:2016wib}
  A.~R.~Brown, L.~Susskind and Y.~Zhao,
  Phys.\ Rev.\ D {\bf 95}, no. 4, 045010 (2017)
  [arXiv:1608.02612 [hep-th]].
  
\bibitem{El-Showk:2014dwa} 
  S.~El-Showk, M.~F.~Paulos, D.~Poland, S.~Rychkov, D.~Simmons-Duffin and A.~Vichi,
  J.\ Stat.\ Phys.\  {\bf 157}, 869 (2014)
  [arXiv:1403.4545 [hep-th]].
  
\bibitem{Kos:2014bka} 
  F.~Kos, D.~Poland and D.~Simmons-Duffin,
  JHEP {\bf 1411}, 109 (2014)
  [arXiv:1406.4858 [hep-th]].
  
\bibitem{Simmons-Duffin:2015qma} 
  D.~Simmons-Duffin,
  JHEP {\bf 1506}, 174 (2015)
  [arXiv:1502.02033 [hep-th]].
  
\bibitem{Kos:2015mba} 
  F.~Kos, D.~Poland, D.~Simmons-Duffin and A.~Vichi,
  JHEP {\bf 1511}, 106 (2015)
  [arXiv:1504.07997 [hep-th]].
  
\bibitem{Iliesiu:2015qra} 
  L.~Iliesiu, F.~Kos, D.~Poland, S.~S.~Pufu, D.~Simmons-Duffin and R.~Yacoby,
  JHEP {\bf 1603}, 120 (2016)
  [arXiv:1508.00012 [hep-th]].

\bibitem{Lin:2015wcg} 
  Y.~H.~Lin, S.~H.~Shao, D.~Simmons-Duffin, Y.~Wang and X.~Yin,
  JHEP {\bf 1705}, 126 (2017)
  [arXiv:1511.04065 [hep-th]].

\bibitem{Poland:2016chs} 
  D.~Poland and D.~Simmons-Duffin,
  Nature Phys.\  {\bf 12}, no. 6, 535 (2016).

\bibitem{Poland:2018epd} 
  D.~Poland, S.~Rychkov and A.~Vichi,
  arXiv:1805.04405 [hep-th].
  
\bibitem{Dymarsky:2017yzx} 
  A.~Dymarsky, F.~Kos, P.~Kravchuk, D.~Poland and D.~Simmons-Duffin,
  JHEP {\bf 1802}, 164 (2018)
  [arXiv:1708.05718 [hep-th]].



\bibitem{FBe}
F.~Brand$\tilde{\text{a}}$o and K.~Svore, Foundations of Computer Science (FOCS), 2017 IEEE 58th Annual Symposium on. IEEE, 2017.  
\bibitem{Ae}
J.~Apeldoorn, A.~Gilyen, S.~Gribling and R.~Wolf, Foundations of Computer Science (FOCS), 2017 IEEE 58th Annual Symposium on. IEEE, 2017.
\bibitem{FB}
F.~Brand$\tilde{\text{a}}$o, A.~Kalev, T.~Li, C.~Lin, K.~Svore and X.~Wu, [arXiv:1710.02581 [quant-ph]].
\bibitem{AA}
J.~Apeldoorn and G.~Andras, [arXiv:1804.05058 [quant-ph]].
\bibitem{sam}
D.~Poulin and P.~Wocjan, Physical review letters, 2009, 103(22): 220502.
\bibitem{AK}
S. Arora and S. Kale. Proceedings of the thirty-ninth annual ACM symposium on Theory of computing. ACM, 2007.

\bibitem{BKLS}
N.~Bao, F.~Kos, B.~Lackey and V.~Su, in progress.




\end{thebibliography}
\end{document}